\documentclass[12pt,aps,floatfix,superscriptaddress,showpacs]{revtex4-1}
\pdfoutput=1 %This is for the ArXiv submission only
\usepackage{amsmath,amssymb,eucal}
\usepackage{graphicx}
\usepackage{subcaption}
\usepackage{float}
\begin{document}

\title{Joint distributions of partial and global maxima of a Brownian Bridge}

\author{Olivier B\'{e}nichou}, 
\address{Laboratoire de Physique Th\'{e}orique de la Mati\`{e}re Condens\'{e}e, UPMC,
CNRS UMR 7600, Sorbonne Universit\'{e}s, 4 Place Jussieu, 75252 Paris Cedex 05, France}
\author{P. L. Krapivsky}
\address{Department of Physics, Boston University, Boston, Massachusetts 02215, USA}
\author{Carlos Mej\'ia-Monasterio}
\address{Laboratory of Physical Properties, Technical University of Madrid, Av. Complutense s/n 28040 Madrid, Spain}
\author{Gleb Oshanin}
\address{Laboratoire de Physique Th\'{e}orique de la Mati\`{e}re Condens\'{e}e, UPMC,
CNRS UMR 7600, Sorbonne Universit\'{e}s, 4 Place Jussieu, 75252 Paris Cedex 05, France}

\begin{abstract}
We analyze the joint distributions and temporal correlations between the partial maximum $m$ and the global maximum $M$ achieved by a Brownian Bridge on the subinterval $[0,t_1]$ and on the entire interval $[0,t]$, respectively. We determine three probability distribution functions: The joint distribution $P(m,M)$ of both maxima; the distribution $P(m)$ of the partial maximum; and the distribution $\Pi(G)$ of the gap between the maxima, $G = M-m$. We present exact results for the moments of these distributions and quantify the 
temporal correlations between $m$ and $M$ by calculating the Pearson correlation coefficient.  
\end{abstract}


%\submitto{\jpa}

\maketitle 

\section{Introduction}

\begin{figure}[t]
\includegraphics[width = .55 \textwidth]{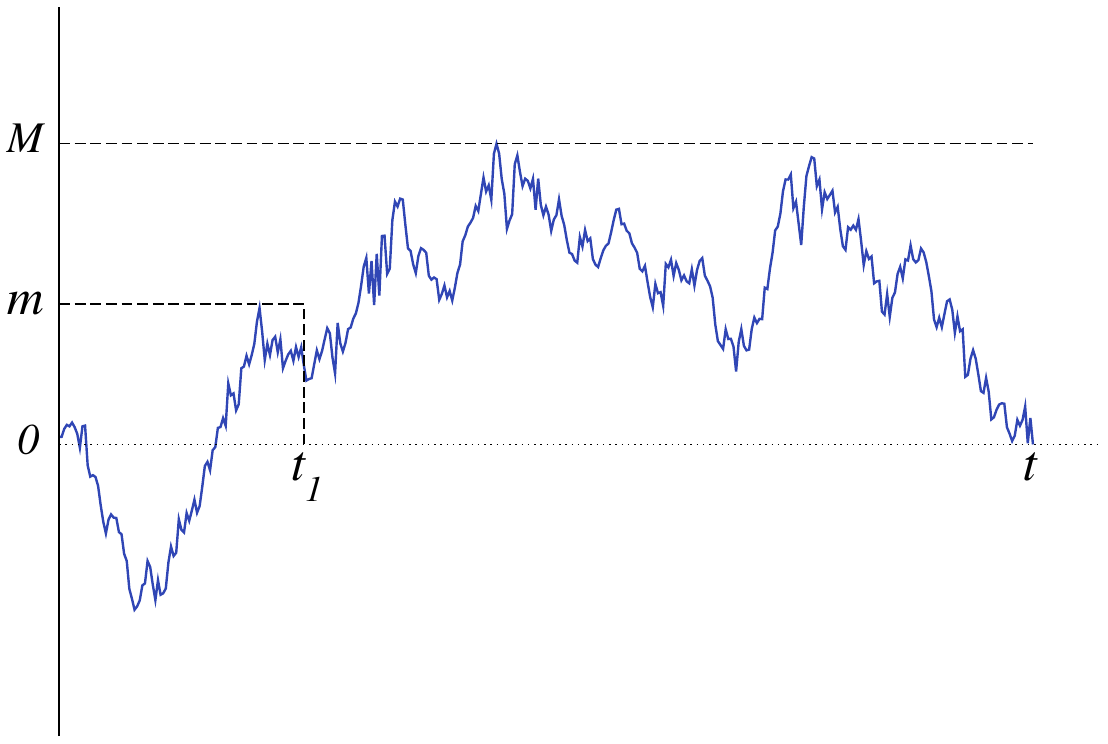}
\caption{A Brownian Bridge (BB) starting at the origin at $s=0$ and returning to the origin at $s=t$.  
The global and partial maxima achieved on the entire interval $[0,t]$ and on the subinterval $[0,t_1]$ with $t_1 \leq t$, respectively, are denoted by $M$ and $m$. In the realization presented on the figure we have $M>m$. The two maxima can also coincide. }
\label{Fig1}
\end{figure}

The Brownian Bridge (BB) is a one-dimensional Brownian motion $B_s$, $0 \leq s \leq t$, which is conditioned to return to the starting point  \cite{peres}. Without loss of generality one can postulate that the BB starts and returns to the origin (see Fig.~\ref{Fig1}): $B_{0} = B_{t} = 0$. BBs admit numerous interpretations. For instance, a BB can be regarded as a stationary $1+1$-dimensional Edwards-Wilkinson interface \cite{ew} in a box with periodic boundary conditions (see, e.g., \cite{sat}).  BBs naturally arise in the analysis of convex hulls of planar Brownian motions \cite{sat2}
and of dephasing due to electron-electron interactions in quasi-$1D$ wires \cite{comtet}, they have been used to model a random potential in studies of diffusion in presence of a strong periodic disorder \cite{dean} and are also relevant for diffusion in disordered non-periodic potentials as they are related to the statistics of transients.
BBs appear in mathematical statistics, e.g., in Kolmogorov-Smirnov tests of the difference between the empirical distributions calculated from a sample and the true distributions governing the sample process \cite{kol,smir,fell,doob} (see also \cite{chiche} for the applications in mathematical finance). BBs are often used in computer science, e.g., in the analysis of the maximal size reached by a dynamic data structure over a long period of time \cite{mari}. In ecology, BBs have been used for an analysis of animal home ranges and migration routes, as well as for estimating the influence of resource selection on movement \cite{horne}. 

Extremal value statistics of the BBs, e.g., statistics of a maximum, a minimum, or a range on the entire time interval $[0,t]$ were studied beginning with classical papers \cite{kol,smir,fell,doob}, and were subsequently generalized for Bessel process (the radius of a $d$-dimensional Brownian motion) with a bridge constraint \cite{gikh,kiefer,pitman}, and also for some conditioned extremal values of BBs \cite{satya3,perret}. The statistics of longest excursions and various non-self-averaging characteristics of BBs have been studied e.g. in Refs.~\cite{laurent95,BD97}. Using a real-space renormalisation group technique, a wealth of results on the extreme value statistics of BBs, reflected Brownian Bridges, Brownian meanders and excursions, as well as more general processes like Bessel Bridges, have been presented in \cite{greg}.    

%\section{Goals and basic notations}

In this paper we investigate the joint statistics and temporal correlations between the partial maximum $m = {\rm max}_{0 \leq s \leq t_1} B_s$ and the global maximum $M = {\rm max}_{0 \leq s \leq t} B_s$ achieved by the BB on the subinterval $[0,t_1]$ and on the entire interval $[0,t]$, see Fig.~\ref{Fig1}. We recently studied similar problems for the unconstrained Brownian motion \cite{we}, and below we compare the outcomes for the BB and the standard Brownian motion (BM) starting at the origin. Due to the ubiquitousness of the BBs, our results admit numerous reformulations. For instance, the analogy with the Edwards-Wilkinson interface asserts that our problem is tantamount to studying the correlations of the maximal height of the interface on the entire interval and the maximal height in a window near one of the fixed boundaries. 

In the next section we determine three probability distribution functions (pdfs): The joint pdf of both maxima, $P(m,M)$; the pdf of the partial maximum, $P(m)$; and the pdf of the gap between the maxima, $\Pi(G)$ with $G = M - m$. Using these distributions, we derive exact expressions for the moments $\mathbb{E}\{m^k\}$ and $\mathbb{E}\{(M-m)^k\}$ with arbitrary $k \geq 0$. We also calculate the Pearson correlation coefficient $\rho(m,M)$ which permits us to quantify the linear correlations between $m$ and $M$.

To determine $P(m,M)$ we use two auxiliary pdfs which describe the BM starting at the origin. One of these quantities is $\Pi_t(m,x)$, the pdf that the BM is at $x$ at time $t$ and it has achieved the maximum $m$ during the time interval $[0,t]$. This pdf is given by (see, e.g., Refs.~\cite{Levy,IM65}) 
\begin{eqnarray}
\label{maxtail}
\Pi_t(m,x) = \dfrac{2 m - x}{2 \sqrt{\pi D^3 t^3}}\, \exp\!\left(- \dfrac{\left(2 m - x\right)^2}{4 D t}\right)\,.
\end{eqnarray}
Another quantity is $S_t(m,x)$, the pdf that the BM does not reach a fixed level $m > 0$ within the time interval $[0,t]$, and appears at position $x$ at time moment $t$.  This survival probability is given by (see, e.g., \cite{red}) 
\begin{eqnarray}
\label{surv}
S_t(m,x) = \dfrac{1}{\sqrt{4 \pi D t}} \left[\exp\!\left(- \dfrac{x^2}{4 D t} \right) - 
\exp\!\left(- \dfrac{(2 m - x)^2}{4 D t} \right)\right]\,.
\end{eqnarray}

\begin{figure}
    \centering
    \begin{subfigure}[b]{0.45\textwidth}
        \includegraphics[width=\textwidth]{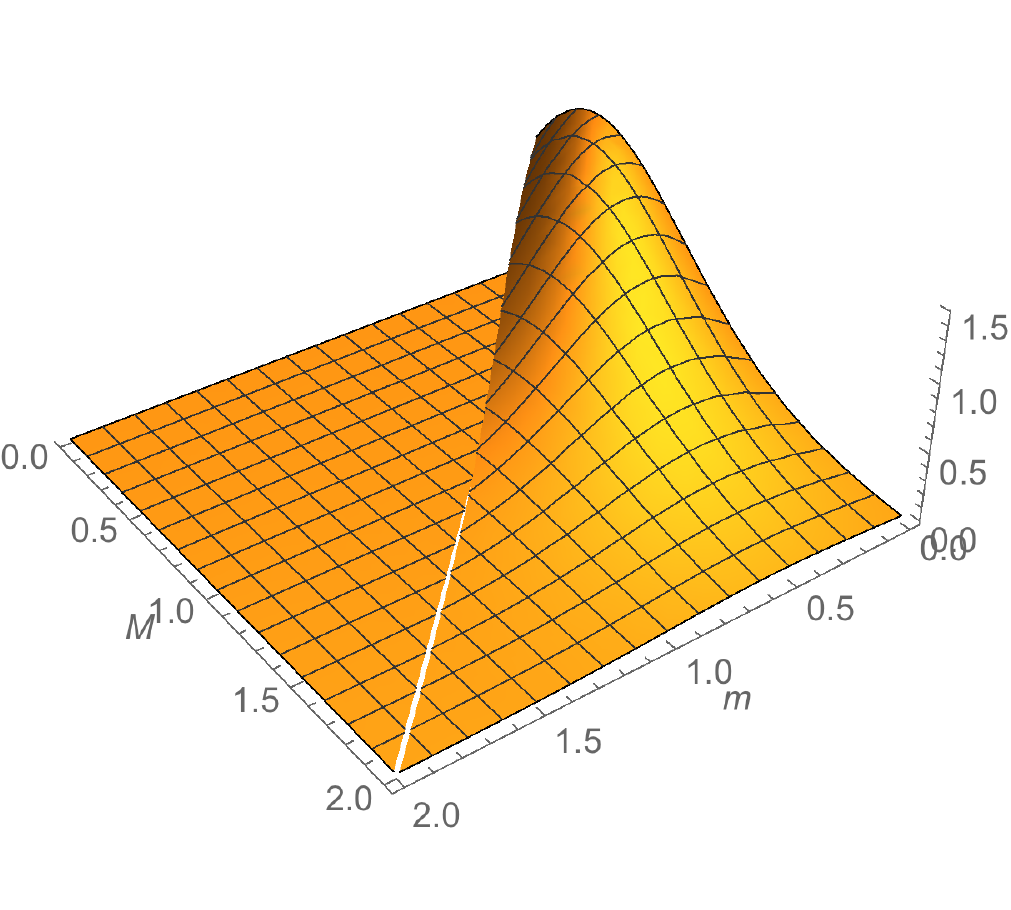}
        \caption{ $z=1/8$}
        \label{distmMz18}
    \end{subfigure}
    \begin{subfigure}[b]{0.45\textwidth}
        \includegraphics[width=\textwidth]{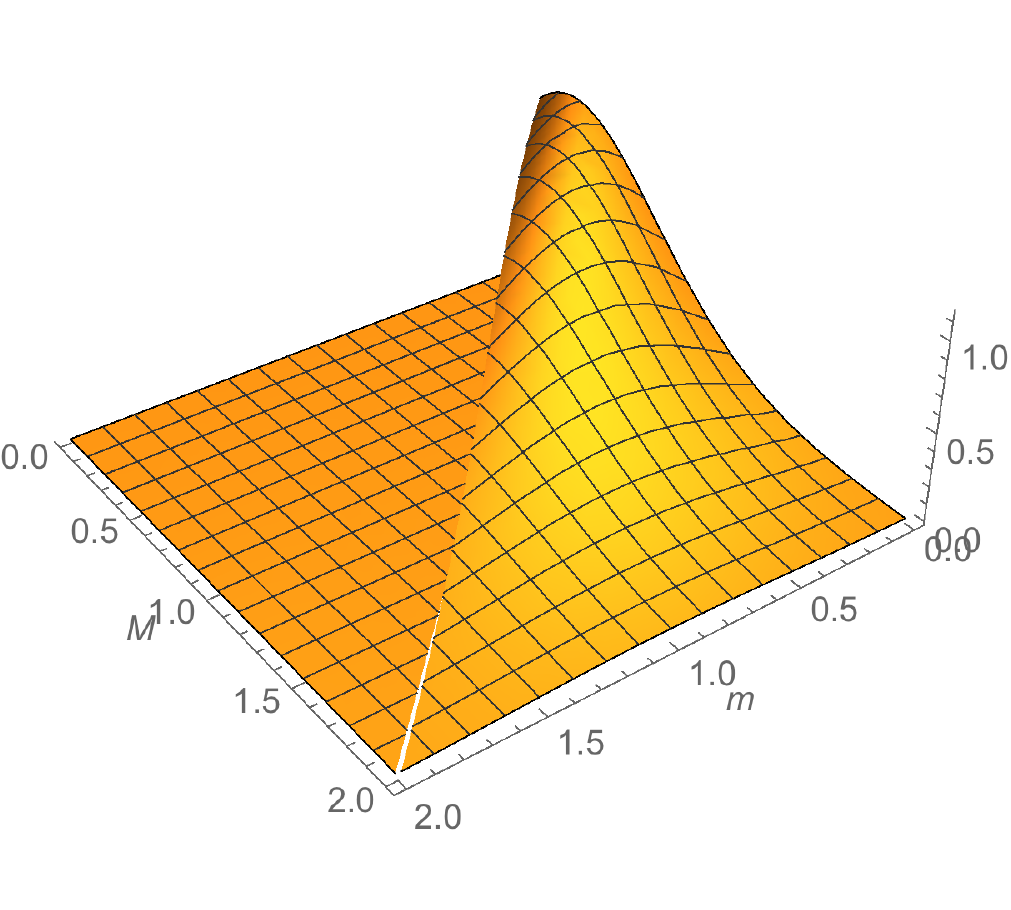}
        \caption{$z=1/4$}
        \label{distmMz14}
    \end{subfigure}
   \begin{subfigure}[b]{0.45\textwidth}
        \includegraphics[width=\textwidth]{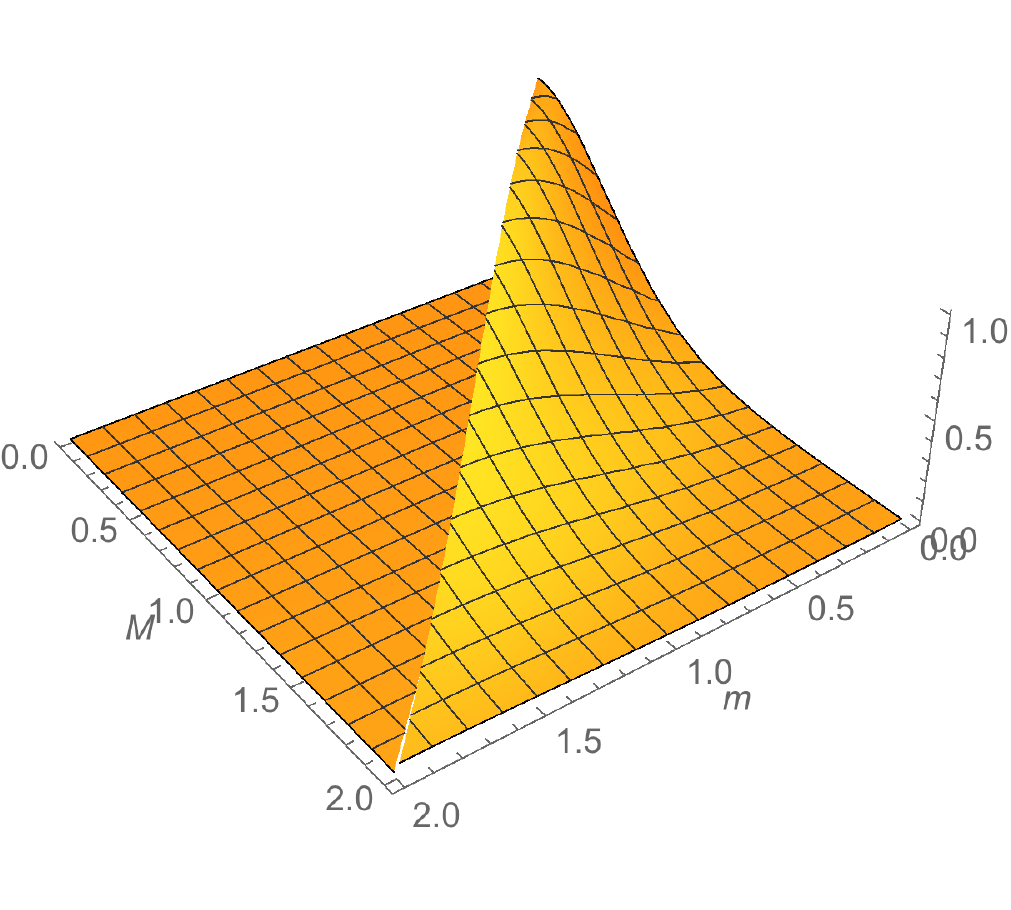}
        \caption{ $z=1/2$}
        \label{distmMz12}
    \end{subfigure}
    \begin{subfigure}[b]{0.45\textwidth}
        \includegraphics[width=\textwidth]{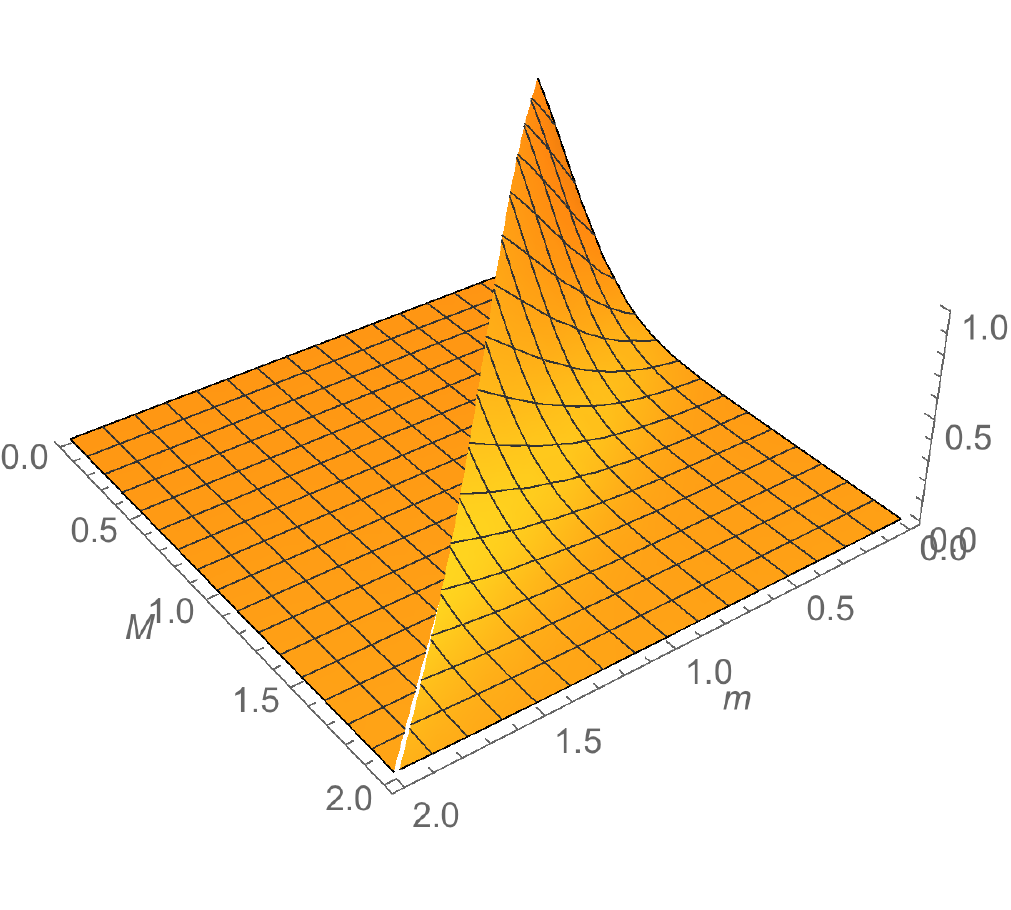}
        \caption{ $z=3/4$}
        \label{distmM34}
    \end{subfigure}
\caption{The joint distribution $P(m,M)$, given by Eq.~\eqref{ppp}, as a function of $m$ and $M$ for several values of $z = t_1/t$. The delta-peak is not shown.
}
\label{pdf}
\end{figure}

%\end{document}

\section{Results}

We compute $P(m,M)$ using the same procedure \cite{we} as for the BM. We denote by $x$ the position of the BB at time $s = t_1$, and we integrate over $-\infty<x<m$ to determine $P(m,M)$. There are two contributions corresponding trajectories with $m=M$ and with $m<M$, the latter occurs e.g. for the trajectory shown in Fig.~\ref{Fig1}. 
Thus we represent $P(m,M)$ as
\begin{eqnarray}
\label{pp}
P(m,M) &=& A \Big(\delta\left(M - m\right)  \int_{-\infty}^m dx \, \Pi_{t_1}\left(m, x\right) \, S_{t - t_1}\left(m,  x\right) + \nonumber\\
 &+& \int^m_{-\infty} dx \, \Pi_{t_1}\left(m, x\right) \, \Pi_{t-t_1}\left(M,x\right) \Big) \,
\end{eqnarray}
The normalization factor $A$ is chosen to ensure that $\int^{\infty}_0 dm \int^{\infty}_{m} dM P(m,M)=1$.
Using \eqref{maxtail} and \eqref{surv} we get $A = 2 \sqrt{\pi D t}$.  

Performing the integrals in \eqref{pp}, we arrive at
\begin{align}
\label{ppp}
&P(m,M) = \dfrac{m}{D t} \exp\left(- \dfrac{m^2}{D t} \right)
\left(1 - {\rm erf}\left(\dfrac{\left(1- 2 z\right) \, m}{2 \sqrt{z (1-z) D t}}\right)\right) \, \delta\left(M - m\right) \nonumber\\
&+ \dfrac{(1-z) \, \left(2 M - m\right) + z \, m}{\sqrt{\pi z (1-z)} \left(D t\right)^{3/2}} \, \exp\!\left(- \dfrac{m^2}{4 D t z} - \dfrac{\left(2 M - m\right)^2}{4 D t (1-z)}\right) \nonumber\\
&-\dfrac{2 \left(M-m\right)^2 - D t}{\left(D t\right)^{2}} \exp\!\left(- \dfrac{\left(M - m\right)^2}{D t}\right) \, {\rm erfc}\!\left(\dfrac{z \left(2 M - m\right) + (1 - z) m}{2 \sqrt{z (1-z) D t}}\right) \,.
\end{align}
Hereinafter we use the shorthand notation $z = t_1/t$.

Equation \eqref{ppp} is the chief result of this paper, it allows us to deduce most of other results. In Fig.~\ref{pdf} we plot $P(m,M)$ [without the delta-peak] for several values of $z$. The pdf $P(m,M)$ is bimodal for $z < 1/2$ (due to the delta-peak for $m=M$) and unimodal for $z \geq 1/2$. In what follows we analyze the characteristic features of the pdf in \eqref{ppp} in more detail.

\subsection{Distribution and moments of the partial maximum}

\begin{figure}
    \centering
    \begin{subfigure}[b]{0.45\textwidth}
        \includegraphics[width=\textwidth]{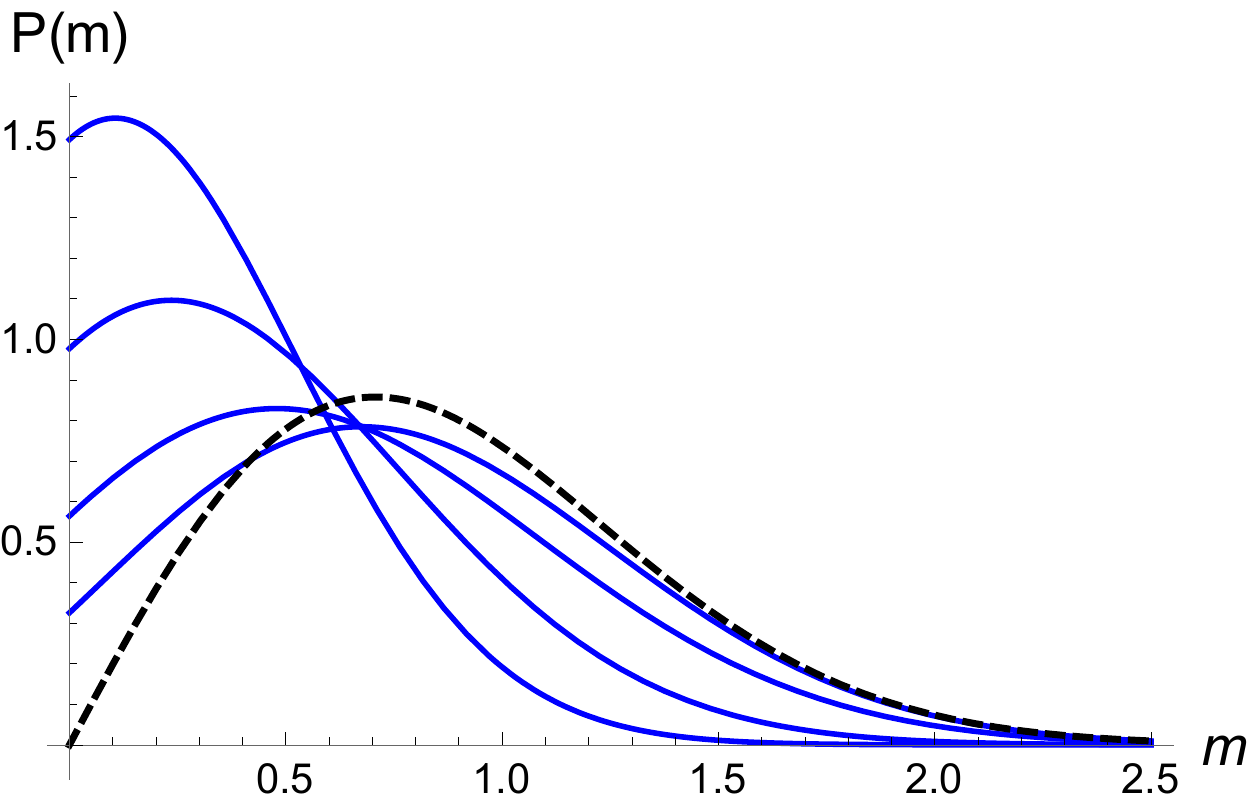}
        \caption{ }
        \label{Fig2}
    \end{subfigure}
     \begin{subfigure}[b]{0.45\textwidth}
        \includegraphics[width=\textwidth]{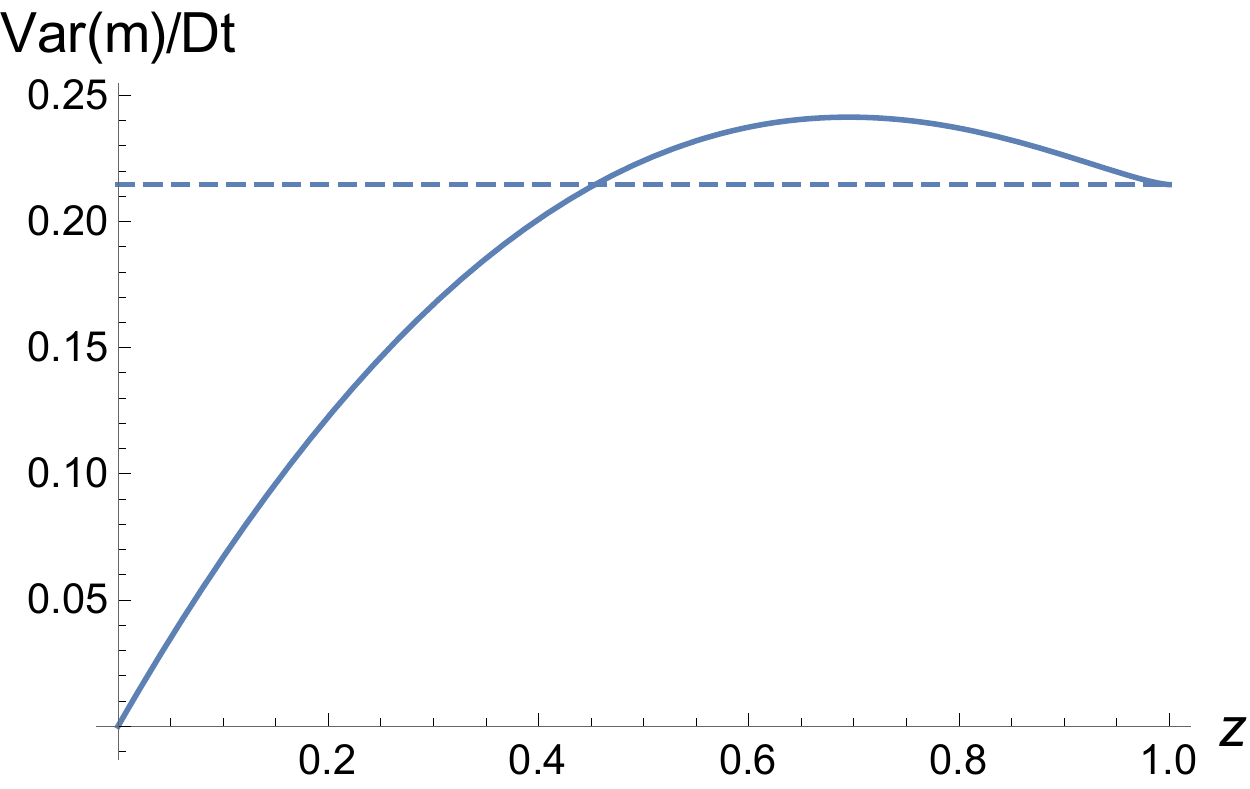}
        \caption{ }
        \label{Fig7}
    \end{subfigure}
\caption{Panel (a): The distribution $P(m)$ of the partial maximum, Eq.~\eqref{m}, vs. $m$ for $D t=1$. The solid curves (top to bottom) correspond to $z=1/8, 1/4, 1/2, 3/4$. The dashed line is the classic result in \eqref{clas}.
Panel (b): The variance ${\rm Var}(m)/D t$ of the partial maximum, Eq.~\eqref{varm}, as a function of $z$ (solid line). The dashed line represents the variance ${\rm Var}(M)/D t \equiv (1- \pi/4)$ of the global maximum $M$. 
}
%\label{pdf}
\end{figure}

We now compute $P(m)$, the distribution of the partial maximum $m$, viz. the maximum of the BB defined on the entire interval $[0,t]$ which is achieved on a subinterval $[0,t_1]$. This quantity can be calculated from \eqref{ppp} by integrating over $M\geq m$:
\begin{eqnarray}
\label{m}
P(m) &=& \int_m^{\infty} dM P(m,M) \nonumber\\
&=& \sqrt{\dfrac{1-z}{\pi z D t}} \exp\left(- \dfrac{m^2}{4 z (1-z) D t}\right) 
\nonumber\\
&+& \dfrac{m}{D t} \exp\left(- \dfrac{m^2}{D t}\right) \left(1 - {\rm erf}\left(\dfrac{1 - 2 z}{2 \sqrt{z (1-z) D t}} \, m\right)\right) \,.
\end{eqnarray}
In the limit $z \to 1$, i.e., when $t_1 \to t$ and $m \to M$,  Eq.~\eqref{m} reduces to the
classic result for the global maximum of the BB (see \cite{kol,smir,fell,doob}):
\begin{align}
\label{clas}
& P(M) = \dfrac{2 M}{D t} \exp\left(- \dfrac{M^2}{D t}\right) \,.
\end{align}
The distributions \eqref{m} and \eqref{clas} are depicted in Fig.~\ref{Fig2}.

%\begin{figure}[t]
%\begin{center}
%\centerline{\includegraphics[width = .5 \textwidth]{Fig2}}
%\caption{The distribution $P(m)$ of the partial maximum, Eq.~\eqref{m}, vs. $m$ for $D t=1$. The solid curves (top to bottom) correspond to $z=1/8, 1/4, 1/2, 3/4$. The dashed line is the classic result in \eqref{clas}.
%\label{Fig2}
%}
%\end{center}
%\end{figure} 
 
\begin{figure}
    \centering
    \begin{subfigure}[b]{0.45\textwidth}
        \includegraphics[width=\textwidth]{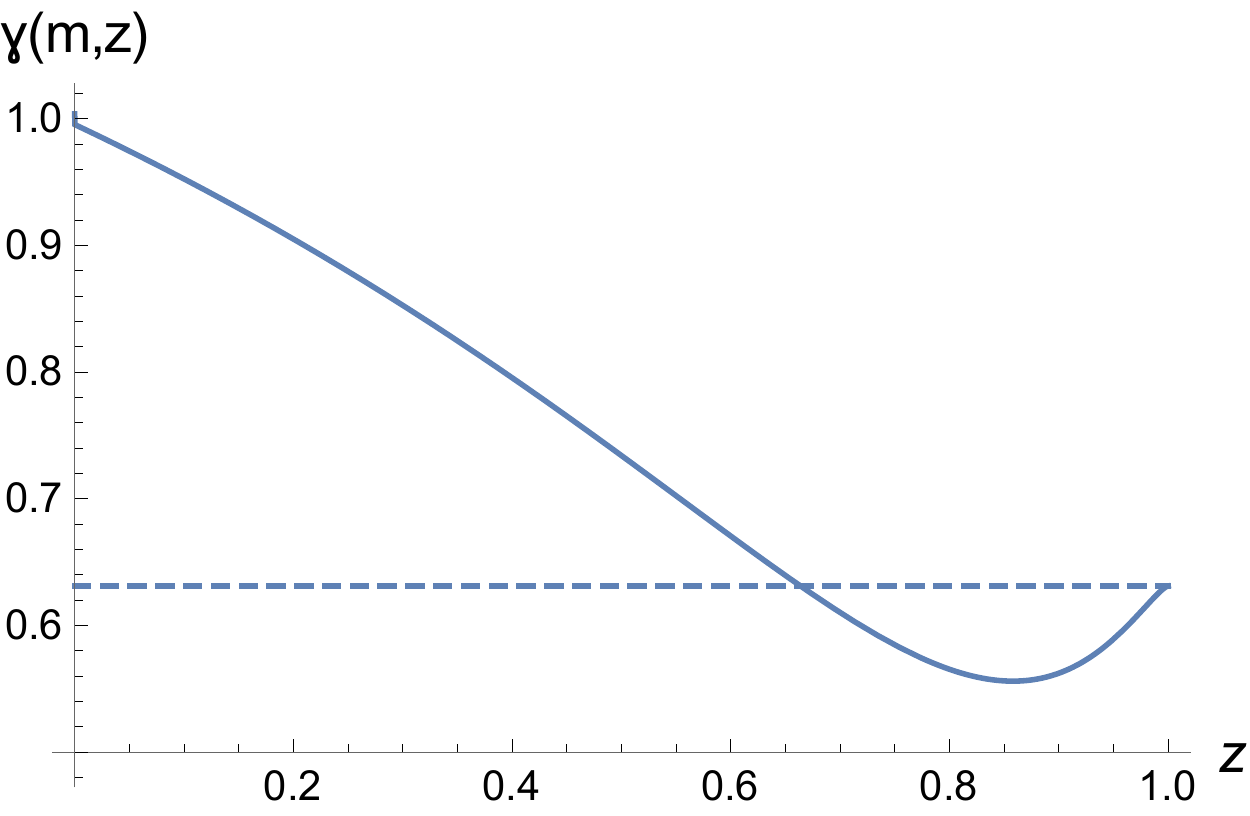}
        \caption{}
        \label{skewness}
    \end{subfigure}
    \begin{subfigure}[b]{0.45\textwidth}
        \includegraphics[width=\textwidth]{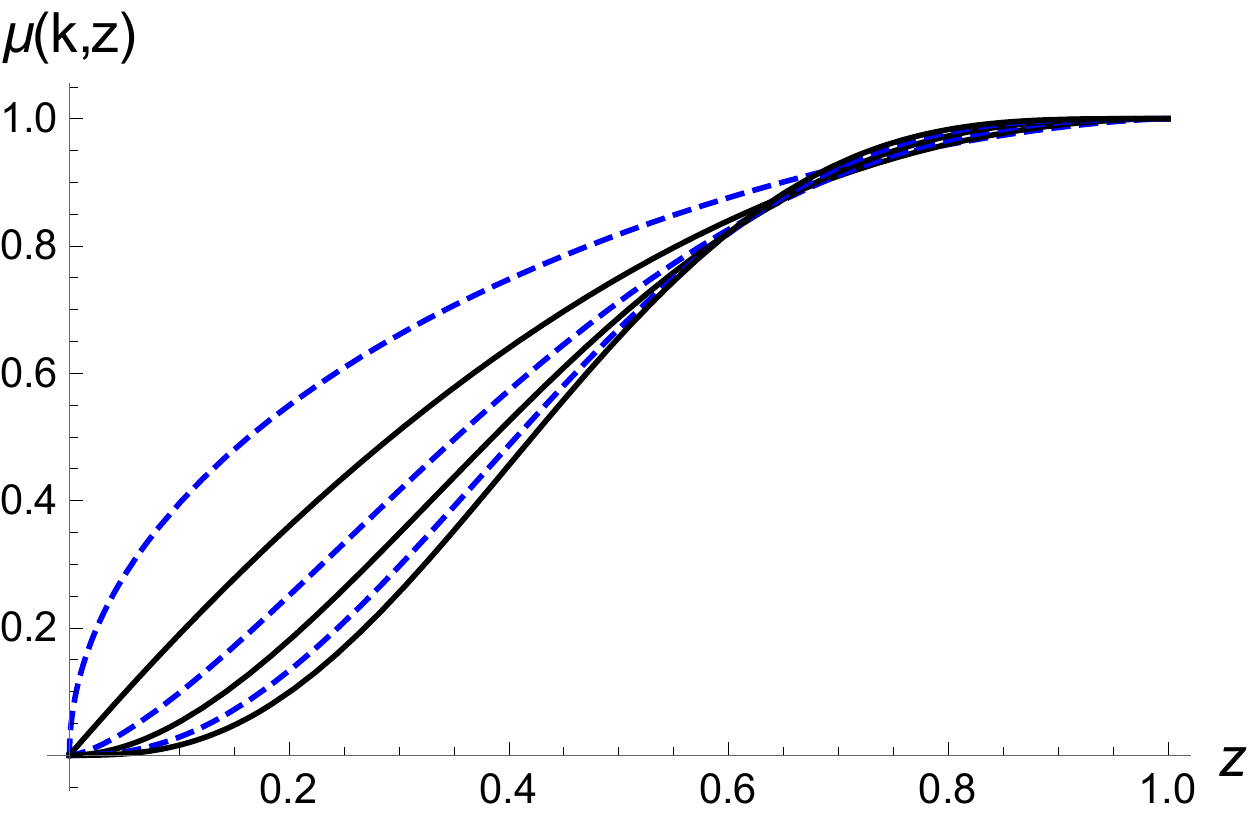}
        \caption{}
        \label{Fig3}
    \end{subfigure}
    \caption{Panel (a): The
    skewness $\gamma(m,z)$ of the pdf of a partial maximum, Eq.~\eqref{skewness1}, as a function of $z$ (solid line). The dashed line represents the skewness $\gamma(M) = 2 (\pi - 3) \sqrt{\pi}/(4 - \pi)^{3/2}$ of the pdf of the global maximum $M$, Eq.~\eqref{clas}. Panel (b): The ratio of the moments $\mu(k,z) = \mathbb{E}\left\{m^k\right\}/\mathbb{E}\left\{M^k\right\}$ vs $z$.
The dashed curves (top to bottom) correspond to $k = 1, 3, 5$. The solid lines (top to bottom) 
correspond to $k=2, 4, 6$.    
}
\label{moments}
\end{figure}

Using \eqref{m} we compute the moments 
\begin{eqnarray}
\label{mom}
\dfrac{\mathbb{E}\left\{m^k\right\}}{\left(D t\right)^{k/2}} &=& \dfrac{1}{2}\,\Gamma\!\left(\dfrac{k+2}{2}\right)  + \dfrac{\left(1-z\right) 2^k\, \Gamma\left(\tfrac{k+1}{2}\right)}{\sqrt{\pi}} \, \left[z \left(1 - z\right)\right]^{k/2} \nonumber\\
&-& \dfrac{1- 2 z}{\sqrt{\pi}}\, \Gamma\left(\dfrac{k+3}{2}\right) \,_2F_1\left(\dfrac{1}{2}, -\dfrac{k}{2}; \dfrac{3}{2}; \left(1- 2 z\right)^2\right)  \,,
\end{eqnarray}
where $\,_2F_1$ is the hypergeometric function. We consider the moments with non-negative integer $k$, although they are well-defined for all $k>-1$. For even integer $k\geq 0$ one can express the moments through the Gegenbauer polynomials, viz.
\begin{eqnarray}
\dfrac{\mathbb{E}\left\{m^k\right\}}{\left(D t\right)^{k/2}} &=& \frac{2^k}{\sqrt{\pi}}\,  \Gamma\!\left(\dfrac{k+1}{2}\right) z^{k/2} \left(1-z\right)^{k/2+1} \nonumber\\
&+& \dfrac{1}{2}\,\Gamma\!\left(\dfrac{k+2}{2}\right) \left(1 + C^{-(k+1)/2}_{k+1}\left(1 - 2 z\right)\right) \,,
\end{eqnarray}
where $C_{\alpha}^{\beta}(x)$ are Gegenbauer polynomials. In particular, for $k=2, 4, 6$
\begin{equation}
\begin{split}
\dfrac{\mathbb{E}\left\{m^2\right\}}{\left(D t\right)}        &= 2 z - z^2 \\
\dfrac{\mathbb{E}\left\{m^4\right\}}{\left(D t\right)^2}    &= 12 z^2 - 16 z^3 + 6 z^4\\
\dfrac{\mathbb{E}\left\{m^6\right\}}{\left(D t\right)^3}    &= 120 z^3 - 270 z^4 + 216 z^5 - 60 z^6
\end{split}
\end{equation}
The moments of the odd order have a more complicated structure and contain the inverse trigonometric function $\arccos(1- 2 z)$. The first three odd moments read
\begin{equation}
\begin{split}
& \dfrac{\mathbb{E}\left\{m\right\}}{\left(D t\right)^{1/2}} = \dfrac{\sqrt{z (1-z)}}{\sqrt{\pi}} + \dfrac{\arccos\left(1- 2 z\right)}{2 \sqrt{\pi}} \\
& \dfrac{\mathbb{E}\left\{m^3\right\}}{\left(D t\right)^{3/2}} = \dfrac{\sqrt{z (1-z)}}{2 \sqrt{\pi}} \left(-3 +14 z - 8 z^2\right) + \dfrac{3 \arccos\left(1- 2 z\right)}{4 \sqrt{\pi}} \\
& \dfrac{\mathbb{E}\left\{m^5\right\}}{\left(D t\right)^{5/2}} = \dfrac{\sqrt{z (1-z)}}{4 \sqrt{\pi}} \left(-15 - 10 z + 248 z^2 - 336 z^3 + 128 z^4\right) + \dfrac{15 \arccos\left(1- 2 z\right)}{8 \sqrt{\pi}} 
\end{split}
\end{equation}
Using explicit expressions for $\mathbb{E}\left\{m\right\}$ and $\mathbb{E}\left\{m^2\right\}$
we determine the variance of the partial maximum: 
\begin{equation}
\label{varm}
\begin{split}
{\rm Var}(m) &= Dt\, V(z)\\
V(z) &=2z-z^2+\dfrac{z^2-z-\sqrt{z (1-z)} \arccos\left(1 - 2 z\right) - \tfrac{1}{4}\arccos^2\left(1 - 2 z\right)}{\pi}\,.
\end{split}
\end{equation}
Interestingly enough, as shown in Fig.~\ref{Fig7}, the variance of $m$ appears to be 
a non-monotonic function of $z$: upon a gradual increase of $z$, ${\rm Var}(m)/Dt$ first grows, crosses 
at $z \approx 0.454$
the dashed line
which defines the corresponding value of the variance of the global maximum M, ${\rm Var}(M)/Dt = (1 - \pi/4)$, attains a maximal value at $z \approx 0.695$ and then decreases reaching finally the level ${\rm Var}(M)/Dt$ at  $z=1$. This is a rather intriguing behavior which shows that in some region the variance of the partial maximum of a BB can be bigger than the variance of the global maximum. Note that the skewness $\gamma(m,z)$ of the pdf $P(m)$ in Eq.~\eqref{m}, defined
as
\begin{equation}
\label{skewness1}
\gamma(m,z) = \dfrac{\kappa_3}{{\rm Var}^{3/2}(m)} \,,
\end{equation}
where $\kappa_3$ is the third cumulant of the pdf $P(m)$, also exhibit a non-monotonic behavior as a function of $z$, see Fig.~\ref{skewness}.

In the limit $z \to 0$, in the leading in $z$ order, we recover from \eqref{mom} the standard expression for the moments of the maximum of an unconstrained Brownian motion on the interval $[0,t_1]$, i.e.,
\begin{align}
\label{momsmall}
& \dfrac{\mathbb{E}\left\{m^k\right\}}{\left(D t_1\right)^{k/2}} \approx \dfrac{2^k \, \Gamma\left(\dfrac{k+1}{2}\right)}{\sqrt{\pi}} \,,
\end{align}
while in the opposite limit $z \to 1$ we have
\begin{align}
\label{mombig}
& \dfrac{\mathbb{E}\left\{M^k\right\}}{\left(D t\right)^{k/2}} \to \Gamma\left(\dfrac{k}{2}+1\right) \,,
\end{align}
which is a standard expression for the moments of the global maximum $M$ of a Brownian Bridge. The moments $\mathbb{E}\left\{m^k\right\}$ as functions of $z$ are plotted in Fig.~\ref{Fig3}.

\subsection{Distribution and moments of the gap between the partial and global maxima}

\begin{figure}[t]
\begin{center}
\centerline{\includegraphics[width = .5 \textwidth]{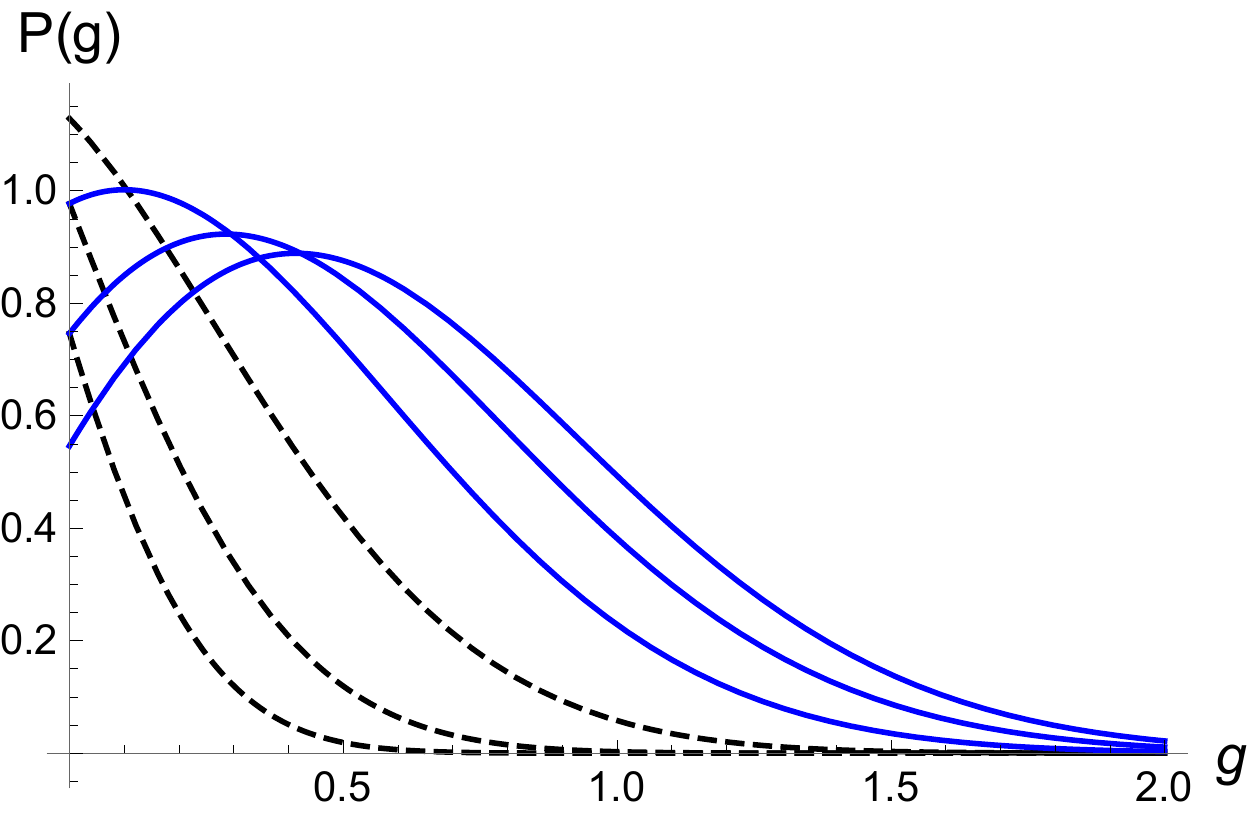}}
\caption{The rescaled gap distribution $\mathcal{P}(g)$, Eq.~\eqref{gap}, vs. rescaled gap $g=G/\sqrt{D t}$ (the delta-peak at $g=0$ is not shown). The dashed curves (top to bottom) correspond to $z=1/2, 3/4, z=7/8$. The solid curves (top to bottom)correspond to $z= 1/4, 1/8, 1/16$.
 \label{Fig4}
}
\end{center}
\end{figure}

From \eqref{ppp}  we derive the distribution $\Pi(G)$ of the gap between $M$ and $m$. Rescaling the gap and the gap distribution
\begin{equation}
\Pi(G) = \frac{1}{\sqrt{Dt}}\,\mathcal{P}(g), \qquad g = \frac{G}{\sqrt{D t}}
\end{equation}
we get 
\begin{eqnarray}
\label{gap}
\mathcal{P}(g) &=& z  \, \delta(g) + 4 \sqrt{\frac{z (1-z)}{\pi}}\,\left(1 -  g^2\right) \exp\!\left(- \dfrac{g^2}{1- z}\right)  \nonumber\\
&+& 2g[2gz + 1- 3 z]\, e^{-g^2}\, {\rm erfc}\left(\sqrt{\dfrac{z}{1-z}}\,g\right)\,.
\end{eqnarray}
This distribution is depicted in Fig.~\ref{Fig4}. For $z \geq 1/2$, the distribution $\mathcal{P}(g)$ is unimodal 
with maximum at $g = 0$. For $z< 1/2$, the distribution is bimodal---in addition to the maximum at $g=0$ (due to the delta-peak) there is a second maximum which moves away from the origin as $z \to 0$.

\begin{figure}
    \centering
    \begin{subfigure}[b]{0.45\textwidth}
        \includegraphics[width=\textwidth]{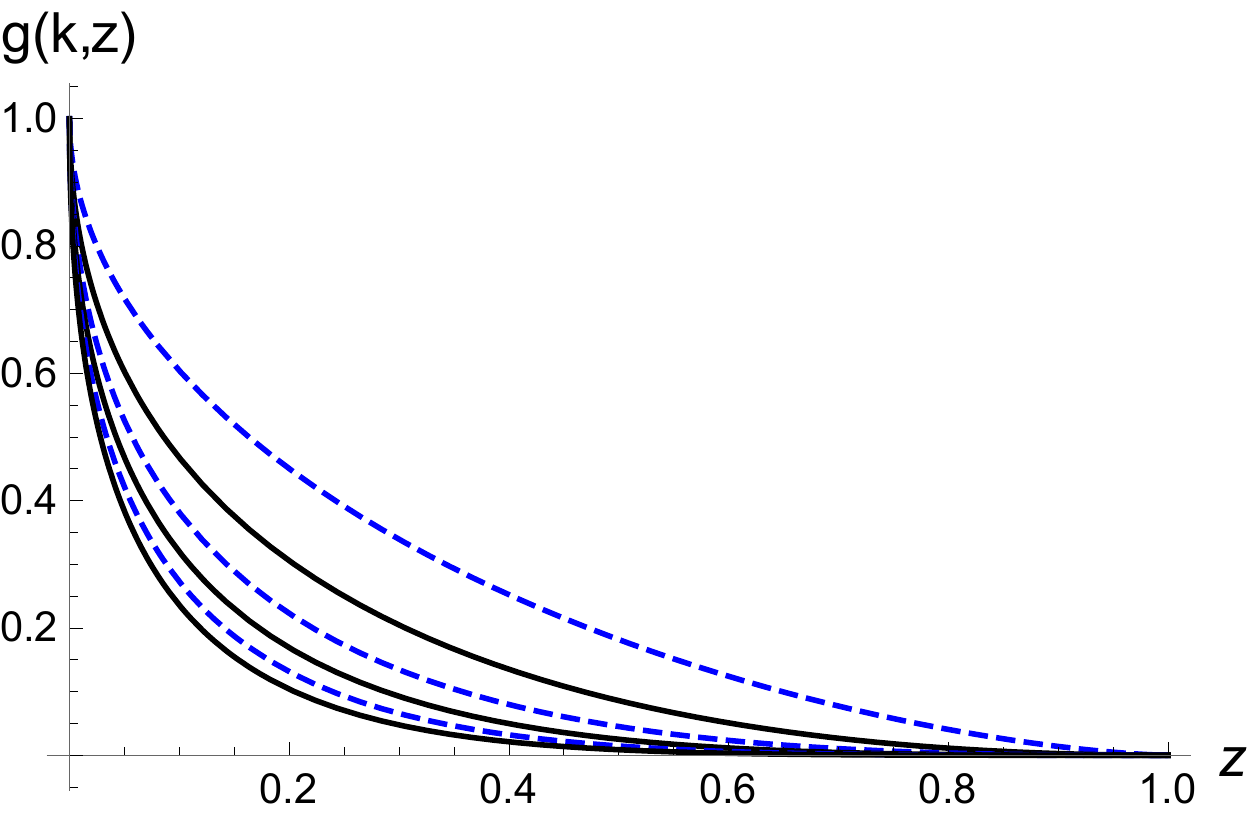}
        \caption{}
        \label{Fig5}
    \end{subfigure}
    \begin{subfigure}[b]{0.45\textwidth}
        \includegraphics[width=\textwidth]{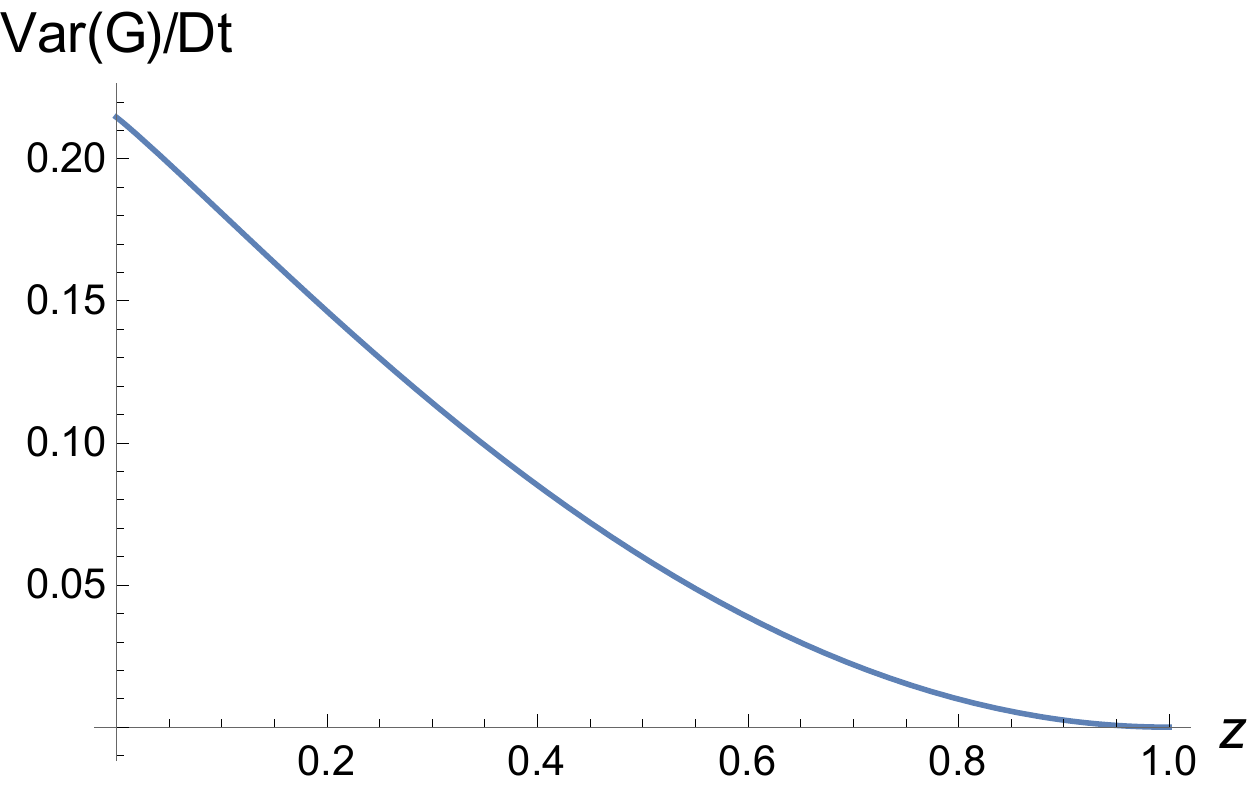}
        \caption{}
        \label{Fig8}
    \end{subfigure}
    \caption{Panel (a): The ratios of the moments $g(k,z)=\mathbb{E}\left\{G^k\right\}/\mathbb{E}\left\{M^k\right\}$ vs. $z$. The dashed curves (top to bottom) correspond to $k=1, 3, 5$. The solid curves (top to bottom)
correspond to $k= 2, 4,6$. Panel (b): The variance ${\rm Var}(G)/D t$ of the gap, Eq.~\eqref{varG}, as a function of $z$.
}
\label{gaps}
\end{figure}

The moments of the gap are found from \eqref{gap} to give
\begin{eqnarray}
\label{momg}
\dfrac{\mathbb{E}\left\{G^k\right\}}{(D t)^{k/2}} & = & 
\dfrac{1}{\sqrt{\pi}}\,\Gamma\left(\dfrac{k+1}{2}\right)  \left(1+ z - (1-z) k\right) \, \sqrt{z} \, \left(1-z\right)^{(k+2)/2} \nonumber\\
&+& \dfrac{\Gamma\left(k+4\right)}{2^{k/2}}\, z (1 - z)^{(k+4)/4} P_{k/2+1}^{-k/2-2}\left(\sqrt{z}\right) \nonumber\\
&+& \dfrac{\Gamma\left(k+2\right)}{2^{k/2}}\, (1 - 3 z) (1- z)^{(k+2)/4} P_{k/2}^{-k/2-1}\left(\sqrt{z}\right) \,,
\end{eqnarray}
where $P_{\alpha}^{\beta}(\cdot)$ are the associated Legendre functions of the first kind. For even $k$ the latter are polynomials, so that the moments of the even order are polynomials of $\sqrt{z}$. For instance
\begin{align}
&\dfrac{\mathbb{E}\left\{G^2\right\}}{(D t)} =  \left(1 - \sqrt{z}\right)^2\,, \nonumber\\
&\dfrac{\mathbb{E}\left\{G^4\right\}}{(D t)^2} = 2 \, \left(1 - \sqrt{z}\right)^3\,, \nonumber\\
&\dfrac{\mathbb{E}\left\{G^6\right\}}{(D t)^3} = \dfrac{3}{2} \left(1 - \sqrt{z}\right)^4 \, \left(4 + \sqrt{z}\right) 
\end{align}
The moments of odd order contain an additional inverse trigonometric function $\arccos(\sqrt{z})$:
\begin{align}
&\dfrac{\mathbb{E}\left\{G\right\}}{(D t)^{1/2}} = \dfrac{1}{\sqrt{\pi}} \left(\arccos\left(\sqrt{z}\right) - \sqrt{z (1-z)}\right) \,, \nonumber\\
&\dfrac{\mathbb{E}\left\{G^3\right\}}{(D t)^{3/2}} =  \dfrac{3}{2 \sqrt{\pi}} \left((1+ 2 z) \arccos\left(\sqrt{z}\right) - 3 \sqrt{z (1 - z)} \right) \nonumber\\
&\dfrac{\mathbb{E}\left\{G^5\right\}}{(D t)^{5/2}} =  \dfrac{5}{4 \sqrt{\pi}} \left(3 (1+ 4 z) \arccos\left(\sqrt{z}\right) - \left(13 + 2 z\right) \sqrt{z (1 - z)} \right) \,.
\end{align}
Using these explicit results one can compute cumulants. For instance, the variance reads
\begin{align}
\label{varG}
\dfrac{{\rm Var}(G)}{D t} = \left(1 - \sqrt{z}\right)^2 - \dfrac{1}{\pi} \left(\arccos\left(\sqrt{z}\right) - \sqrt{z (1-z)}\right)^2
\end{align}
In Fig.~\ref{Fig5} we plot $\mathbb{E}\left\{G^k\right\}$ vs $z$ for several integer 
values of $k$, while Fig.~\ref{Fig8} presents the variance of the gap vs $z$.

\begin{figure}[t]
\includegraphics[width = .5 \textwidth]{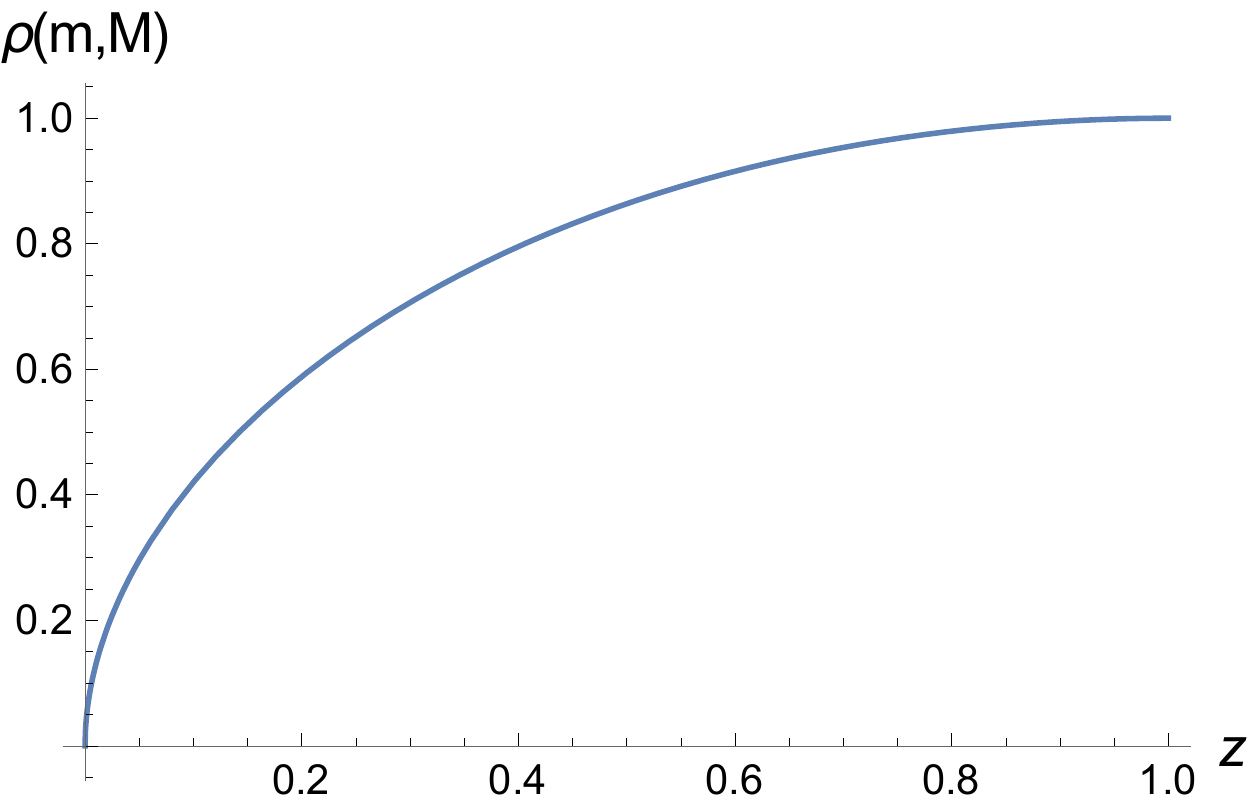}
\caption{Pearson correlation coefficient $\rho(m,M)$, \eqref{pearson}, versus $z$.}
\label{Fig10}
\end{figure} 
 
\subsection{Two-time correlations between the partial and global maxima}

Let us determine the Pearson correlation coefficient of partial $m$ and global $M$ maxima. By definition
\begin{align}
\label{pearson}
&\rho\left(m,M\right) = \dfrac{\mathbb{E}\left\{m M \right\} - \mathbb{E}\left\{m\right\} \mathbb{E}\left\{M\right\}}{\sqrt{{\rm Var}(m) \, {\rm Var}\left(M\right)}} \,.
\end{align}
We have already computed all terms in Eq.~\eqref{pearson} apart from the cross-moment of two maxima  $\mathbb{E}\left\{m M \right\}$. This cross-moment can be determined from \eqref{ppp} to give
\begin{align}
& \mathbb{E}\left\{m M\right\} = \dfrac{Dt}{2} \left(2 \sqrt{z} + z - z^2\right) 
\end{align}
leading to
\begin{equation}
\label{RmM}
\rho\left(m,M\right) = \dfrac{2 \sqrt{z} + z - z^2 - \sqrt{z (1-z)} - \tfrac{1}{2}\arccos\left(1 - 2 z\right)}{\sqrt{(4-\pi)V(z)}}
\end{equation}
with $V(z)$ defined in Eq.~\eqref{varm}. 

In Fig.~\ref{Fig10} we plot the Pearson's coefficient as a function of $z$. The Pearson coefficient approaches unity, $\rho(m,M) \to 1$, when $z\to 1$, i.e. $t_1 \to t$.  Indeed, $m$ and $M$ are almost completely correlated in this region. 
The more precise asymptotic behavior is  
\begin{align}
\rho(m,M) = 1 - \dfrac{4 \sqrt{1-z}}{4-\pi} + O\left(\left(1-z\right)^{3/2}\right) \,.
\end{align}
Conversely, $\rho(m,M) \to 0$ when $z \to 0$ implying that $m$ and $M$ become uncorrelated. More precisely, one gets
\begin{align}
\rho(m,M) = \sqrt{\dfrac{\pi}{2 (\pi - 2)(4 - \pi)}}\, \sqrt{z} + O\left(z\right) \,,
\end{align} 
implying that correlations vanish slowly, $\rho(m,M)  \sim \sqrt{t_1/t}$.

\section{Conclusions}

We have determined the joint statistics and temporal correlations between a partial and global extremes of one-dimensional Brownian bridges. We have calculated the joint probability distribution function of two maxima,  
the pdf of the partial maximum and the pdf of the gap $G = M - m$. We also derived exact expressions for the moments $\mathbb{E}\{m^k\}$ and $\mathbb{E}\{(M-m)^k\}$ with arbitrary $k \geq 0$ and computed the Pearson correlation coefficient $\rho(m,M)$ quantifying the correlations between $m$ and $M$. Our results for the one-dimensional Brownian bridges can be generalized to the general Bessel process---the radius of $d$-dimensional Brownian motion, with the bridge constraint. The calculations are very similar, one should use explicit expressions for $\Pi_t(m,x)$ obtained in \cite{greg}.

\vspace{0.2in}
The research of O.B. was supported by ERC grant FPTOpt-277998.
C.M-M acknowledges the support from the Spanish MICINN grants MTM2012-39101-C02-01 and MTM2015-63914-P.

%\section*{References}

\end{document}